\documentclass[amsmath,amssymb,pra,aps,showpacs,superscriptaddress,twocolumn]{revtex4-2}
\usepackage{amsmath,amsfonts,amssymb,amsthm,graphics,graphicx,epsfig,bbm}
\usepackage[colorlinks=true,citecolor=blue,linkcolor=red,urlcolor=blue]{hyperref}
\usepackage[usenames]{color}
\usepackage{graphicx}
\usepackage{subfigure}
\usepackage{amsmath}
\usepackage{epsfig}
\usepackage{dcolumn}
\usepackage{bm}
\usepackage{color}
\usepackage{epstopdf}
\usepackage{amssymb}
\usepackage{amstext}
\usepackage{latexsym}
\usepackage{hyperref}
\usepackage{amsfonts}
\usepackage{psfrag}
\usepackage{xcolor}
\usepackage[normalem]{ulem}
\usepackage{dsfont}
\usepackage{txfonts}
\usepackage{overpic}
\usepackage{ifthen}
\usepackage{natbib}

\usepackage[capitalise]{cleveref}

\newcommand{\an}[2]{\ifthenelse{\equal{#1}{}}{\ensuremath{\hat{#1}_{#2}}}{\ensuremath{\hat{#1}^{\protect\phantom{\dagger}}_{#2}}}}

\newcommand{\UNAL}{Departamento de Matemáticas, Universidad Nacional de Colombia, Medellín, Antioquia, 050034, Colombia}

\newcommand{\GEO}{McDonough School of Business, Georgetown University, 37th and O Streets, Washington DC, 20057, US}

\newcommand{\ALI}{Team Foods Colombia S.A., KR. 11 \# 84 - 09, piso 4, Bogotá DC, 110221, Colombia}

\newcommand{\FIS}{Departamento de Física, Universidad Nacional de Colombia, Medellín, Antioquia, 050034, Colombia}

\begin{document}

\title{Quantum Ecosystem Research and Analysis in Colombia}

\author{Cristian E. Bello}
\email{crbellor@unal.edu.co} 
\affiliation{\UNAL}

\author{Benjamin Harper}
\email{bph46@georgetown.edu} 
\affiliation{\GEO}

\author{Camilo A. Castro}
\affiliation{\ALI}

\author{Alcides Montoya C.}
\affiliation{\FIS}

\date{\today}

\begin{abstract}
The rapid growth of quantum computing in the last few years has led many countries to make relevant public investments in that field. However, Colombia lacks any investment or legislation in this area. Most previous studies have focused on other countries, contributing to the region's significant lag. In this paper, we propose efforts to include quantum computing as a fundamental pillar of Colombia's development plan. In this research we involved three stakeholders: academia (Universidad Nacional de Colombia), industry (Alianza Team), and government (Vice Ministry of Digital Transformation). We anticipate that our work will provide a connection between all stakeholders involved in public investments in quantum technology in the country and will facilitate its legislation.
\end{abstract}

\maketitle

\section{Introduction}
Quantum computing has the potential to transform the world on a scale comparable to that of the printing press, the internal combustion engine, or the internet. This emerging technology, in which bits of information, known as \textit{qubits}, can simultaneously exist in multiple states, promises to create algorithms capable of solving complex problems at speeds incomparable with a high speed up compared to classical computers. Field such as healthcare, pharmaceutical research, cryptography, national defense, transportation, and energy, can benefit from the speed and precision offered by quantum algorithms. However, the transformative impact of this technology has been unevenly distributed. With few exceptions, such as public funding for certain practical applications, the development of quantum technology is concentrated primarily in the Global North. The highly industralization, energy-independent, and predominantly social-democratic nations of North America, Europe, East and South Asia, and the Pacific region are leading this technological revolution \cite{qureca}.

These disparities have motivated the present project. If quantum technology is would be a catalyst for economic growth and social transformation on a global scale, the opportunities for funding and business development in this field should not be limited by geography or gross domestic product (GDP) \cite{imf}. This project focuses on Latin America, a region characterized by emerging economic growth driven by academic disciplines in science, technology, engineering, and mathematics (STEM) \cite{STEM}, but which has seen limited direct investment and governmental awareness of the impending technological revolution. By identifying quantum computing efforts and developing awareness strategies in Colombia, we seek to ensure that the transformative opportunities offered by this technology, much like qubits, can overlap and be available everywhere, benefiting everyone simultaneously.

\section{Background}
\begin{figure*}[hbpt!]
\centering
\includegraphics[width=\textwidth]{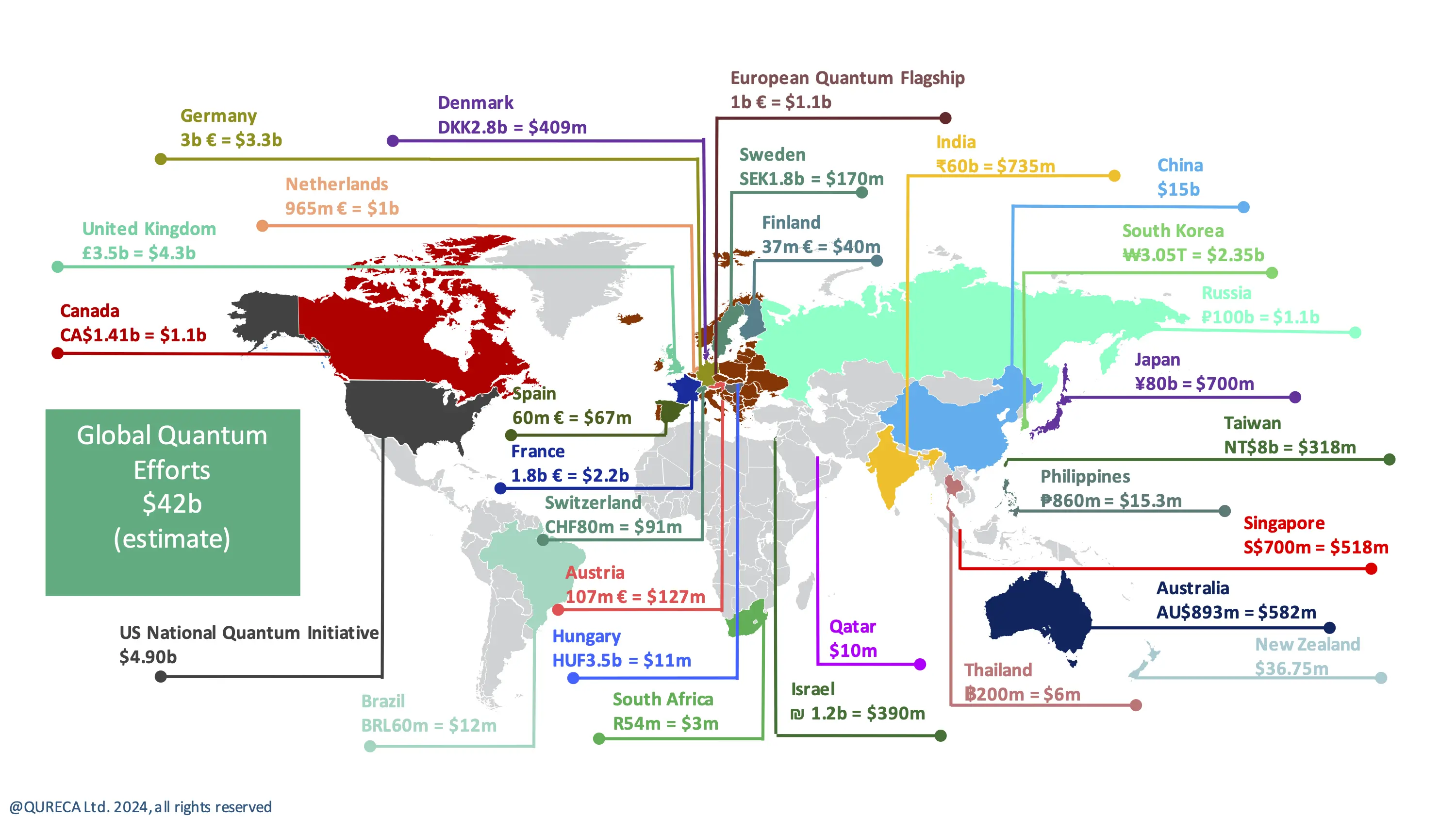}
\caption{Quantum Initiatives Worldwide $2024$. Taken from: \url{https://www.qureca.com}.}
\label{fig:qureca}
\end{figure*}

This project emerged as a direct result of the infographic "Global Quantum Ecosystems - An Overview of Quantum Initiatives Worldwide" \ref{fig:qureca}, created by \textit{Dr. Araceli Venegas-Gómez}, General Director of Qureca, a Spain-based company that provides consulting services for the development of quantum workforce, primarily in Europe \cite{QURECAWebsite}. Although the chart highlights the considerable investments by nation-states, most in billions in quantum research, the vast majority of these countries are in Europe, North America, and East Asia \ref{fig:qureca}. Only exception of Brazil \cite{BRICS}, where Qureca forecasts a $\$12$ billion investment in quantum research by $2030$, the infographic representation of Latin America south of Mexico in $2023$ was entirely gray: no significant governmental investment was recorded. Our hypothesis is that while Qureca's information is accurate and useful, it may not represent a complete picture. After all, the absence of national investment in quantum does not necessarily imply a lack of regional or private quantum research or use case applications. Further investigation revealed a fascinating core of regional quantum initiatives around Peru and Uruguay.

Peru hosts the annual Quantum Latino conference, where practical and research advancements are shared with scientists and academics from across the region \cite{QuantumLatino}. Uruguay, on the other hand, boasts a successful company called Quantum South, Quantum South, that uses quantum software to optimize air cargo planning and goods transportation. This company emerged from an academic research group at the University of Montevideo in $2019$ \cite{QuantumSouth}. These initial findings fueled our optimism: there was significant professional interest in quantum growth in Latin America, as well as a successful business application. From this point, our focus specifically turned to the country that would serve as a testing lab for our research and recommendations: Colombia.

Secondary sources provided little information on the state of quantum computing in Colombia. National planning documents focus on national digital transformation; if private companies or universities were researching quantum computing, it was not widely publicized. However, the overall potential for technological growth in Colombia is really promising when applying the Political, Economic, Social, and Technological (PEST) research framework to the country. Politically, President \textit{Gustavo Francisco Petro Urrego} recently published the National Development Plan $2022 - 2026$, where digitalization and reindustrialization of the Colombian economy are key aspects, with a shift away from raw material exports towards a knowledge-based economy, technology, and sustainable industry \cite{dnp}. There is a high political interest for transformative change. Economically, Colombia has a surprisingly strong Information and Communication Technology (ICT) market. Colombian exports of software and hardware, designed by Colombian IT professionals, programmers, and data engineers, surged during the pandemic and have not declined. According to the World Bank, Colombian ICT exports reached $1.04$ billion in $2023$ \cite{WorldBank}, with a projected total market value for $2024$ (combining foreign and domestic sales) of $11.3$ billion, demonstrating a solid year-over-year growth rate of $6.6\%$ over the past five years \cite{tradingeconomics}.

It is clear that Colombia has a solid base of STEM professionals and demonstrated economic growth upon which to build quantum research and application. Socially, we found a long-standing national commitment within the Colombian scientific community. The \textit{Red Caldas de Científicos e Ingenieros Colombianos en el Exterior}, created in the $1990$ \cite{Sabios}, allows Colombian STEM professionals studying or working abroad to network, share international project opportunities, and ultimately motivate each other to return and work in science and technology in their native country \cite{Grans1996}. From a technological perspective, although it is unlikely that Colombia will ever host quantum computing hardware (due to its mostly tropical and steppe climate, as well as significant issues with energy and internet supply), its infrastructure is conducive to increasing the use of quantum software programs. The Ministry of ICT has established wireless internet connections in $48.5$ million households, and $5G$ technology is expected to be available nationwide by $2025$ \cite{mintic}. With a promising analysis, we established connections with potential Colombian clients. Drawing on Professor \textit{Mike Ryan} research on public-private academic interdependencies \cite{MichaelRyan}, to understand the current state of quantum research and funding in Colombia, we first needed to assess how advanced each of these sectors was in their respective quantum efforts, as well as the potential challenges that could be addressed through increased interdependence. To apply the model, we selected stakeholders from academia (Universidad Nacional de Colombia), industry (Alianza Team), and government (Vice Ministry of Digital Transformation).

\section{Research Findings}
Field research was conducted in person in Bogotá, Colombia. Methodologies included reviewing secondary informational materials on the missions and capabilities of stakeholders. Most of our findings were developed during face-to-face interviews, which influenced our recommendations and, in at least one case, directly contributed to the development of business and partnership opportunities. Overall, our findings revealed that, although nascent, there is a small but healthy quantum ecosystem of companies and universities that would greatly benefit from a unified quantum public policy.

As previously mentioned, Alianza Team is a company specializing in vegetable fats and oils, with 75 years of experience in the food industry. The company operates in Colombia, Mexico, Chile and, since $2022$, in the \textit{Raleigh-Durham-Chapel Hill} research triangle in North Carolina, USA. Despite recent work in applied research with artificial intelligence, Alianza found it necessary to venture into quantum software due to the inflationary environment, driven by both necessity and a curiosity for innovation. In this way, Alianza, in collaboration with its longstanding partner IBM, leveraged quantum computing software to enhance its product portfolio. Thanks to quantum software, Alianza was able to optimize the use and stability of triglycerides in margarine and shortening production, designing a superior product with a texture identical to conventional products and an optimal melting point for use in bakery products. While classical artificial intelligence algorithms took $45$ minutes to aid the design of margarines, Quantum computing, combined with Alianza's expertise in fat product design methodologies, reduced this time to just $3$ seconds. Alianza currently has an additional project underway utilizing quantum machine learning algorithms for the development of tools that reduce time and meet the needs of product developers. However, Alianza's biggest challenge is client perception. The company’s clients are unfamiliar with the validity of quantum-designed food products and may hesitate to incorporate them into their portfolios due to the lack of quantum regulation in Colombia. Currently, there is no national policy regulating or managing the use of quantum technology in manufacturing, which is especially challenging in Colombia's food industry, already subject to strict regulations. Until there is a comprehensive public policy on quantum technology, Alianza issue could be summarized as: emerging quantum applications without buyers.

In the academic sphere, the Quantum Computing Program at the Universidad Nacional de Colombia (UNAL) comprises approximately $55$ professors and students, primarily from Colombia but also from other parts of the countries. Actually, this team has conducted theoretical research in material science and quantum machine learning, although it is beginning to explore practical applications of quantum technology. UNAL short-term goals include establishing partnerships with government and private industry to fund their research.

In the long term, UNAL aims to retain and expand Colombia's quantum workforce, helping students find well-paying jobs in computing that allow them to stay in the country throughout their careers. The Quantum Computing Program is the only one of its kind in Colombia and is gaining popularity, especially online, where program members weekly teach quantum computing courses on platforms like Twitch\cite{Twitch} and YouTube\cite{YouTube}, catching tens of thousands of views. However, while program members are passionate, it is challenging to expand the knowledge base necessary for continuous growth. Currently, UNAL offers only one quantum computing course at the graduate level. If Alianza challenge can be described as applications without clients, UNAL could be summarized as research without applications.

Regarding the Vice Ministry of Digital Transformation, there are currently no government policy documents defining quantum priorities or regulating the use of quantum technology in industry, government, or academia. The Vice Ministry informed us about the steps necessary to add quantum technology to Colombia’s national priorities, particularly through Law $152$, the Organic Law of the National Development Plan. This law outlines the concrete steps for adding new priorities to Colombia's National Development Plan. Similar to the United States, strategic advocacy groups and unions have a significant influence on policy formulation added to the National Development Plan, which is published every four years. The most influential organization in this process is the National Business Association of Colombia (ANDI). A key document for our findings is the CONPES, a policy that, once enacted into law, facilitates coordination among various government sectors to achieve a defined goal. The Vice Ministry has several strengths that could support the development of quantum public policies. First, Colombia recently joined the Organization for Economic Cooperation and Development (OECD). To maintain the benefits of membership, including foreign direct investment opportunities with other member countries, Colombia must align its national policies with those of the OECD. Although there is currently no national policy or governance on quantum computing, the OECD offers extensive guidance on the development, risks, and benefits of quantum technology \cite{oecd}. The Colombian government has access to existing OECD frameworks to develop national quantum public policies.

\begin{figure}[htbp!]
\centering
\includegraphics[width=1\linewidth]{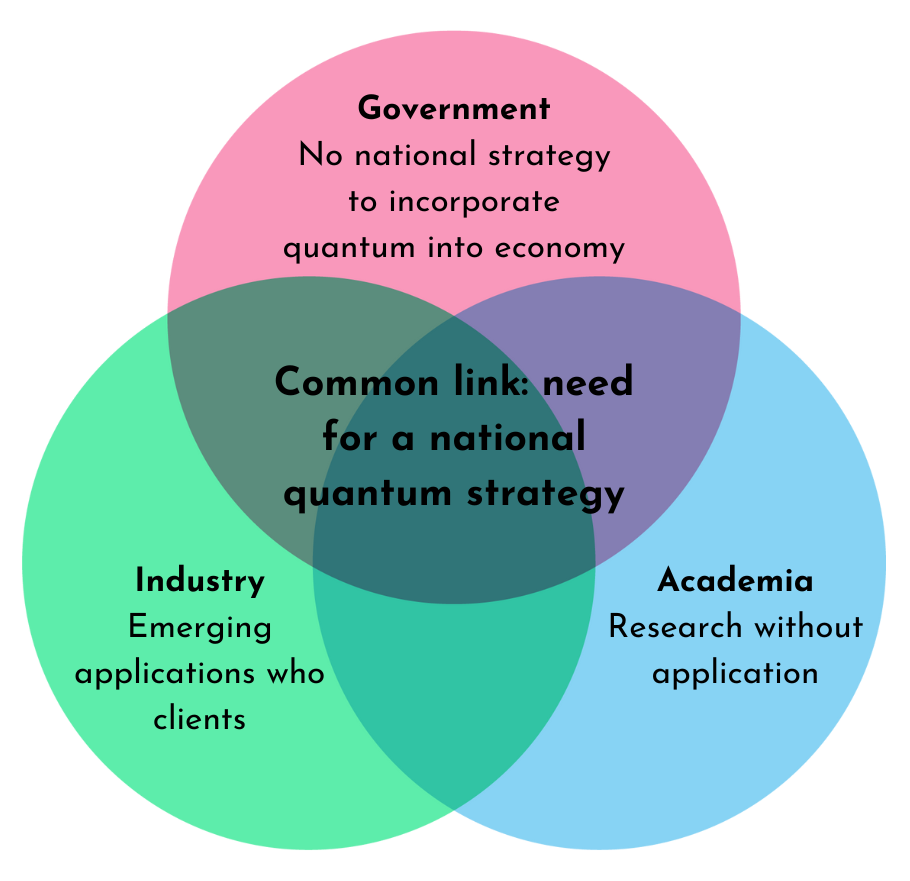}
\caption{Key Finding on Interdependencies.}
\label{fig:key}
\end{figure}

However, since there is no quantum public policy even in the development phase, there is no skateholders network (other ministries, industry unions, presidential advisors, etc.) that could initiate a debate on the future of quantum technology in Colombia. Many of the current policies in the National Development Plan are focused on immediate concerns (reducing inequality, decreasing dependence on oil). This led us to question: Why should the government prioritize research and investment in a technology whose real horizon is between five and ten years in the future? The field research on Alianza Team’s quantum algorithms, UNAL’s emerging efforts, and the Vice Ministry’s public policy has revealed what Ryan calls the \textit{shared interdependencies} \ref{fig:key} of the public, private, and academic sectors \cite{MichaelRyan}.

\section{Strategic Recommendations}
\begin{figure}[htbp!]
    \centering
    \includegraphics[width=1\linewidth]{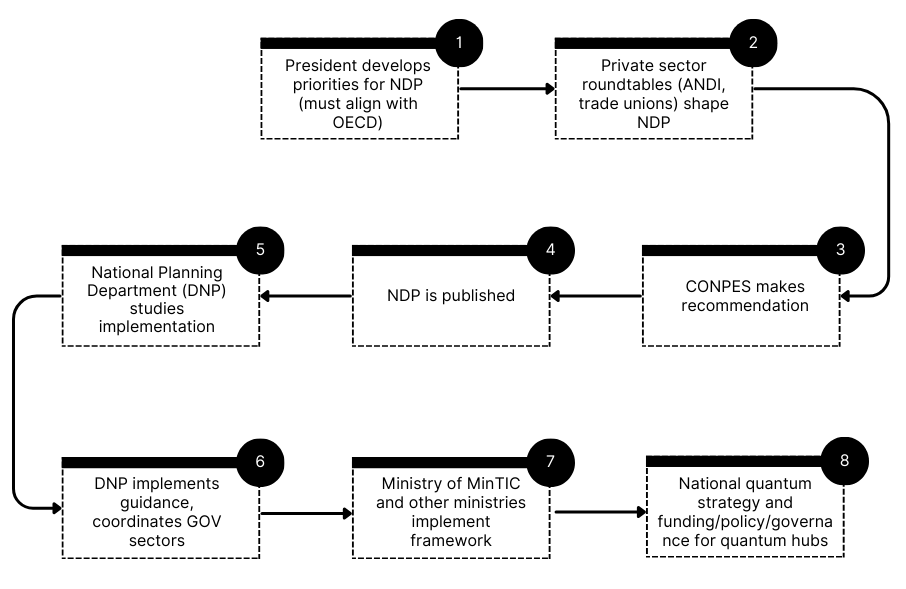}
    \caption{Pathway for Quantum Public Policy.}
    \label{fig:public-policy}
\end{figure}

The stakeholders developed a some recommendations, starting with identifying potential strategies and eliminating those that did not offer a concise solution to the lack of a unified national quantum strategy, as identified during the research phase. We developed a deep understanding of the Colombian legislative process, specifically how new priorities are added to the National Development Plan, how they are codified, funded, and implemented \ref{fig:public-policy}. Once a priority is approved, it is aligned and integrated into the National Development Plan, which is published every four years in conjunction with the start of a new presidential term.

The current National Development Plan was published in $2022$, making $2026$ a key date for the objectives of our campaign. From then on, the National Planning Department (DNP) coordinates the development and actionable milestones with the various interested ministries. In our case, it is likely that the DNP will delegate the responsibility for quantum development to the Ministry of Information and Communication Technologies (MinTIC) and the Ministry of Science, Technology, and Innovation. Our proposed final state includes direct national investment in quantum research groups (such as UNAL), companies using quantum technology (such as Alianza), and the establishment of quantum development centers. We identified two crucial initial stages in policy development: the President's meeting with the National Association of Entrepreneurs of Colombia (ANDI) and the publication of the CONPES as key strategic awareness opportunities.

Once we understood the process for quantum policy to become law, our next step was to identify other stakeholders who, although not part of our field research, would play a strategic role in the public law sphere. Additional stakeholders include the Ministry of Science, Technology, and Innovation and the Mission of Wise Men, a lobbying group formed by leading Colombian figures in STEM advocating, among other things, for an annual investment of $1.2\%$ of Colombia’s GDP in science, technology, and medicine \cite{pavas}. Our next goal was to align existing and emerging quantum research efforts in Colombia with the country’s two most important national policies: the OECD and the National Development Plan. This stage proved relatively straightforward, as the nascent quantum research at UNAL and Alianza fits perfectly with the OECD and PND goals for Colombia’s development. The OECD directs Colombia to prioritize innovation in ICT, and the President’s PND further instructs the government to increase the use and optimization of data as a Digital Public Good, as vital as roads or the education system. Alianza and UNAL are already contributing to these priorities by using quantum algorithms to expand vast data libraries and enhance sustainable food production, food security, and traffic patterns in Colombia's largest city.

The OECD also requires Colombia to promote economic growth, and the PND prioritizes education and workforce retraining in response to change. Models for such change already exist at UNAL and Alianza, which are training and hiring the next generation of digitally literate Colombian professionals. When quantum computing reaches a global scale (as AI did in $2023$), Colombia may be prepared with a digital workforce with quantum knowledge to navigate the change. The OECD calls on Colombia to improve the quality of its government services, while the PND goes a step further, calling for broad reindustrialization and increased national competitiveness and innovation. UNAL and Alianza, already familiar with quantum learning algorithms and their real-world applicable results, can lead the way. With MinTIC, the DNP, and other key players, quantum computing centers, or "hubs," could be established to drive this development. After aligning our stakeholders and campaign with national and OECD priorities, we focused on developing a global framework for implementing achievable quantum goals.

The foundation of our framework is our own strategy: telling the story of current efforts in Colombia and the opportunities that exist to fund and develop this transformative technology. The pillars of the framework, the tactics that comprise our strategy, include four concrete objectives for the national government:

\begin{itemize}
\item Develop a CONPES on quantum investment and public policy, and add an action plan for the development of quantum technology in Colombia.
\item Include a paragraph on quantum in the National Development Plan $2026 - 2030$.
\item Collaborate with ANDI to help companies and universities exploring quantum technology network and share research opportunities, or integrate existing efforts.
\item Make a significant yet realistic investment of $10.4$ million USD ($1.2\%$ of Colombia’s ICT GDP), as recommended by the Mission of Wise Men, in quantum research through UNAL, Colombia's most important public university.
\end{itemize}

Finally, our ultimate goal, which serves as the highest objective of our strategy, is for quantum research, applications, and public policy to be codified in the PND $2026 - 2030$. ANDI, the DNP, MinTIC, and other stakeholders will work together to fund and promote new and existing quantum research and applications.

\section{Conclusion}
At the conclusion of the field research and the beginning of the working groups for strategy development, the Universidad Nacional de Colombia (UNAL) and Alianza Team held a series of meetings. The outcome was highly rewarding: what initially started as an initiative to assess each party's capabilities evolved into the exploration of concrete projects with potential industrial applications. As a direct result of this project, Alianza and UNAL are exploring possibilities for collaboration, including training Alianza's engineers using the same algorithms developed by Alianza and analyzing the company's quantum data. In exchange, Alianza has provided UNAL with opportunities to connect with other companies using quantum technology in Colombia, including IBM. Most notably, this relationship may facilitate the integration of UNAL students into the Colombian workforce upon graduation, ensuring that Colombian STEM talent remains in the country and contributes to what CONPES $3975$ has aptly termed the ``New Industrial Revolution'' \cite{conpes}. Through thorough field research and a recommendation strategy based on raising awareness of the potential of quantum technology, we firmly believe that as the world progresses toward the next major technological transformation, Colombia is well-positioned to lead the way.

\section{Acknowledgments}
C.E.-B. and  A.M.-C. acknowledgs funding from the project “Centro de excelencia en computación cuántica e inteligencia artificial”, HERMES code $52893$, UNAL.
\section{Declarations}

\textbf{Competing Interests:} The authors declare the following competing interests: C.A.-C. was employed by Team Foods Colombia S.A., during the development of this work.

\appendix


\bibliography{main}

\end{document}